\documentclass[10pt,journal,a4paper]{IEEEtran} 

\usepackage{amssymb}
\usepackage{amsmath}
\usepackage{amsmath,bm}
\usepackage{amsthm}
\usepackage{graphicx}
\usepackage{subfigure}
\usepackage{cite}
\usepackage{enumerate}
\usepackage{algorithm}
\usepackage{algorithmicx}  
\usepackage{algpseudocode} 
\usepackage{color, soul}
\usepackage{stfloats}
\allowdisplaybreaks[4]
\begin{document}

\newtheorem{definition}{\bf ~~Definition}
\newtheorem{observation}{\bf ~~Observation}
\newtheorem{theorem}{\bf ~~Theorem}
\newtheorem{proposition}{\bf ~~Proposition}
\newtheorem{remark}{\bf ~~Remark}

\renewcommand{\algorithmicrequire}{\textbf{Input:}} 
\renewcommand{\algorithmicensure}{\textbf{Output:}} 

\title{\Large{Reconfigurable Intelligent Surface assisted Multi-user Communications:\\ How Many Reflective Elements Do We Need?}}
\author{
{Hongliang Zhang}, \IEEEmembership{Member, IEEE},
{Boya Di}, \IEEEmembership{Member, IEEE}, 
{Zhu Han}, \IEEEmembership{Fellow, IEEE},\\
{H. Vincent Poor}, \IEEEmembership{Life Fellow, IEEE},
{and Lingyang Song}, \IEEEmembership{Fellow, IEEE}

\thanks{H. Zhang is with Department of Electronics, Peking University, Beijing, China, and also with Department of Electrical Engineering, Princeton University, NJ, USA (Email: hongliang.zhang92@gmail.com).}

\thanks{B. Di is with Department of Electronics, Peking University, Beijing, China, and also with Department of Computing, Imperial College of London, London, UK  (Email: diboya92@gmail.com).}

\thanks{Z. Han is with Electrical and Computer Engineering Department, University of Houston, Houston, TX, USA, and also with the Department of Computer Science and Engineering, Kyung Hee University, Seoul, South Korea (Email: zhuhan22@gmail.com).}

\thanks{H. V. Poor Department of Electrical Engineering, Princeton University, NJ, USA (Email: poor@princeton.edu).}

\thanks{L. Song is with Department of Electronics, Peking University, Beijing, China  (Email: lingyang.song@pku.edu.cn).}

\vspace{-6mm}}

\maketitle

\begin{abstract}
Reconfigurable intelligent surfaces~(RISs) consisting of multiple reflective elements are a promising technique to enhance communication quality as they can create favorable propagation conditions. In this letter, we characterize the fundamental relations between the number of the reflective elements and the system sum-rate in RIS-assisted multi-user communications. It is known from previous works that the received signal-to-noise ratio~(SNR) can linearly increase with the squared number of RIS reflective elements, but how many elements are sufficient to provide an acceptable system sum-rate still remains an open problem. To this end, we derive the asymptotic capacity with zero-forcing (ZF) precoding, and then discuss how many reflective elements are required so that the ratio of the system sum-rate to the capacity can exceed a predefined threshold. Numerical results verify our analysis.
\end{abstract}
\vspace{-2mm}
\section{Introduction}
Reconfigurable intelligent surfaces~(RISs) have emerged as a cost-effective solution to tackle the data surge in future wireless systems~\cite{M-2019}. By controlling the electromagnetic response (e.g. phase shifts) for a large number of low-cost elements, an RIS can provide a more favorable propagation environment between transmitters and receivers, thus improving the link quality~\cite{MHLKZG}.

In the literature, most existing works focus on the phase shift optimization/analysis in RIS assisted wireless communications. 
In \cite{BHLYZH}, the authors proposed a hybrid beamforming scheme for a multi-user RIS assisted multi-input single-output~(MISO) system together with a phase shift optimization algorithm to maximize the system sum-rate. The authors in \cite{YCZDYJ-2020} optimized the power allocation at the base station~(BS) as well as phase shifts at the RIS to maximize the data rate. The authors in \cite{HBLZ-2020} analyzed the effect of the limited phase shifts on the system data rate. As suggested in \cite{HBLZ-2020}, the size of the RIS will also influence the system sum-rate. However, \emph{how many RIS reflective elements are sufficient to provide an acceptable system sum-rate} remains an open problem in the literature.

In this letter, we consider an RIS assisted downlink multi-user cellular network. To quantify the impact of the number of RIS reflective elements on the system sum-rate, we first provide an asymptotic analysis of the system capacity for the RIS-assisted downlink multi-user MISO communications with zero-forcing~(ZF) precoding which is easy to implement. Based on this capacity analysis, we further discuss how many RIS reflective elements are sufficient to provide an acceptable system sum-rate.  

The rest of this paper is organized as follows. In Section \ref{System}, we introduce a system model for RIS assisted multi-user communications. In Section \ref{Rate}, the asymptotic achievable data rate is derived. The relation between the number of RIS reflective elements and the system sum-rate is discussed in Section \ref{analysis}. Numerical results in Section \ref{simulation} validate our analysis. Finally, conclusions are drawn in Section \ref{sec:conclusion}.

\emph{Notation:} Boldface lower and upper case symbols denote vectors and matrices, respectively. $\|\cdot\|_F$ represents the Frobenius norm. $\mathbb{C}^{M \times N}$ denotes a complex matrix with dimensions $M \times N$. Conjugate transpose and matrix inverse operators are denoted by $(\cdot)^{H}$ and $(\cdot)^{-1}$, respectively. We use $\bm{I}_K$ to denote the $K \times K$ identity matrix. $\mathbb{E}[\cdot]$, $\text{Tr}\{\cdot\}$, and $\det(\cdot)$ are expectation, trace, and determinant operators, respectively. We also use $[\bm{A}]_{k,m}$ to represent the element in the $k$-th row and the $m$-th column of a matrix $\bm{A}$.

\vspace{-2mm}
\section{System Model}%
\label{System}

\begin{figure}[!t]
	\centering
	\includegraphics[width=3.0in]{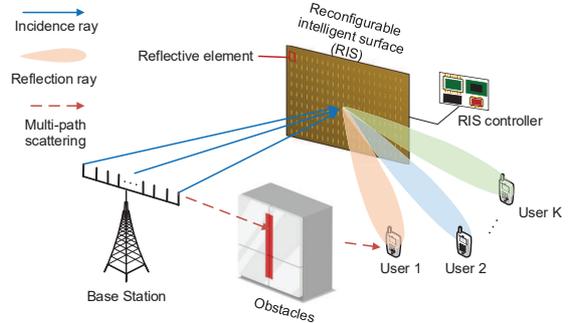}
	\vspace{-2mm}
	\caption{System model of RIS assisted downlink multi-user MISO networks.}
	\vspace{-4mm}
	\label{scenario}
\end{figure}

\subsection{Scenario Description}
Consider a narrow-band downlink multi-user MISO network as shown in Fig.~\ref{scenario}, where one BS with $M$ antennas serves $K$ single-antenna users, where $M \geq K$. Due to the dynamic wireless environment involving unexpected fading and potential obstacles, the Light-of-Sight~(LoS) link between the cellular users and the BS may not be stable or even falls into a complete outage\footnote{The RIS is more likely to be deployed in the cases where the LoS link does not exist to further extend the coverage.}. To enhance the QoS of the communication link, an RIS is adopted to reflect the signal from the BS and directly project the signals to the users by actively shaping the propagation environment into a desirable form. 

The RIS is composed of $N$ electrically controlled RIS reflective elements with the size of $a \times b$. Each reflective element can adjust the phase shift by leveraging positive-intrinsic-negative~(PIN) diodes. A PIN diode can be switched between ``ON" and ``OFF" states, based on which the metal plate can add a different phase shift to the reflected signal. Define $\theta_{n}$ as the phase shift for reflective element $n$,
and the reflection factor of reflective element $n$ can be written by $\Gamma_{n} = \Gamma e^{-j \theta_{n}}$, where the reflection amplitude $\Gamma \in [0,1]$ is a constant.

\vspace{-3mm}
\subsection{Channel Model}

\begin{figure}[!t]
	\centering
	\includegraphics[width=3.0in]{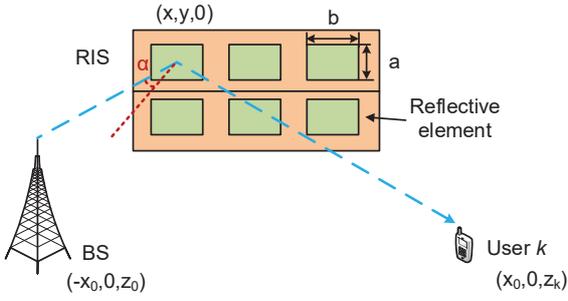}
	\vspace{-2mm}
	\caption{Channel model on the RIS assisted downlink communications.}
	\vspace{-4mm}
	\label{illustration}
\end{figure}

Let $\bm{G} \in \mathbb{C}^{K \times M}$ be the channel matrix between the BS and users, where $g^{km} \triangleq [\bm{G}]_{km}$ is the channel coefficient between user $k$ and antenna $m$ at the BS. Since each reflective element will reflect the signals from the BS to the users, the channel $g_{km}$ consists of $N$ paths, and we denote the channel gain from antenna $m$ to user $k$ through the $n$-th reflective element as $g^{km}_n$. 

The channel matrix $\bm{G}$ models independent fast fading, path loss, and RIS response. To be specific, $g_{n}^{km}$ can be written by
\begin{equation}
g_{n}^{km} = (\beta_{n}^{k})^{-1/2}h_{n}^{km}\Gamma_{n}.
\end{equation}
Here, $h_{n}^{km}$ is independent fast fading coefficient with $\mathbb{E}[h_{n}^{km}] = 0$ and $\mathbb{E}[(h_{n}^{km})^2] = 1$. Based on the results in~\cite{OEE-2020}, the path loss is related with the incident angle $\alpha_n^k$, i.e., $\beta_{n}^{k} = \frac{64\pi^3}{Aab\lambda^2}\frac{(l_n d_n^k)^2}{\cos^3(\alpha_n^k)}$, where $A$ is the antenna gain, $\lambda$ is the wavelength, $l_n$ is the distance between the BS and the $n$-th reflective element\footnote{We assume that the distance between the BS and the RIS is much larger than the margin of two antennas, and thus the distances between the RIS and different antennas are assumed to be the same.}, and $d_n^k$ is the distance between user $k$ and the $n$-the reflective element. Therefore, we have 
\begin{equation}
g^{km} = \sum\limits_{n = 1}^N g_n^{km} = (\bar{\beta}^{k})^{-1/2}\sum\limits_{n = 1}^N h_{n}^{km}\Gamma_{n},
\end{equation}
where $\bar{\beta}^{k}$ is the equivalent path loss between the BS and user $k$ through the RIS.
Therefore, channel matrix $\bm{G}$ can be written by
\begin{equation}
\bm{G} = \bm{B}^{-1/2}\bm{H},
\end{equation} 
where $\bm{B}$ is a diagonal matrix whose $k$-th diagonal reflective element being $\bar{\beta}^k$ and the reflective element at the $k$-th column and the $m$-th row of $\bm{H}$ is equal to $\sum\limits_{n = 1}^N h_{n}^{km}\Gamma_{n}$.

\subsection{Achievable Rate with Zero-forcing Precoding}

To serve $K$ users simultaneously, the BS first encodes the data symbols for different users with a normalized precoding matrix $\bm{W} \in \mathbb{C}^{M \times K}$. In this letter, we adopt the ZF precoder as it can obtain a near-optimal solution with a low complexity~\cite{BHLYZH}. At the BS, the signals are assumed to be transmitted under a normalized power allocation matrix $\bm{\Lambda}$ which satisfies $\text{Tr}\left\{\bm{\Lambda}\bm{\Lambda}^H\right\} = 1$. Denote the intended signal vector for $K$ users as $\bm{s}$ satisfying $\mathbb{E}[\bm{s}\bm{s}^H] = \frac{P}{K}\bm{I}_K$, where $P$ is the transmitted power. Therefore, the transmitted signals at the BS can be given by $\bm{x} = \bm{W\Lambda s}$. Due to ZF precoding, after the reflection by the RIS, the received signal at the users can be written by
\begin{equation}
\bm{y} = \bm{G} \bm{W\Lambda s} + \bm{n},
\end{equation} 
where the $k$-th reflective element of $\bm{y}$ is the received signal for user $k$ and $\bm{n} \sim \mathcal{CN}(\bm{0},\sigma^2)$ is the additive white Gaussian noise. Assume that the channel information is available for the transmitter. For ZF precoding, the $k$-th column of $\bm{W}$ can be written by
\begin{equation}
	\bm{w}_k = \frac{\bm{v}_k}{\|\bm{v}_k\|_F},
\end{equation}
where $\bm{v}_k$ is the $k$-th column of matrix $\bm{V}$ where
\begin{equation}
\bm{V} = \bm{G}^{H}\left(\bm{G}\bm{G}^{H}\right)^{-1}.
\end{equation}

Consequently, with the ZF procoding, the signal-to-noise ratio~(SNR) for user $k$ can be expressed as
\begin{equation}\label{sinr}
	\gamma_k = \frac{P\Lambda_k}{K\sigma^2}\left(\bm{g}_{k}\bm{w}_k\right)\left(\bm{g}_{k}\bm{w}_k\right)^{H},
\end{equation}
where $\bm{g}_k$ is the $k$-th row of $\bm{G}$ and $\Lambda_k = \left[\bm{\Lambda}\bm{\Lambda}^{H}\right]_{k,k}$.
Here, we assume that the channel state information is perfectly obtained by the BS, and thus $\bm{\Lambda}$ is assumed to be a constant in the following which can be obtained by water-filling algorithm~\cite{BHLYZH}.

Use above notations, we can express the data rate for user $k$ as
\begin{equation}
r_k = \mathbb{E}\left[\log_2\left(1 + \gamma_k\right)\right].
\end{equation}

\section{Analysis on Asymptotic Capacity}\label{Rate}

According to \cite{DP-2005}, the system capacity can be achieved by dirty paper coding, where
\begin{equation}
C = \mathbb{E}\left[\sum\limits_{k}\log_2\left(1 + \frac{P \Lambda_k}{K\sigma^2}[\bm{G}\bm{G}^H]_{k,k}\right)\right].
\end{equation}

In the following, we will provide a proposition to show the upper bound of the system capacity.
Before this position, we first give a proposition on the channel matrix.

\begin{proposition}\label{pro_central}
	When the number of reflective element $N$ is large and the phase shifts of the RIS are given, i.e., $\theta_n$ is constant, we have\footnote{It is worthwhile to point out that this proposition holds with any distribution of $h_n^{km}$.}
	\begin{equation}\label{central}
	\frac{\sum\limits_{n = 1}^{N} h_n^{km}\Gamma_{n}}{\Gamma \sqrt{N}} \sim  \mathcal{CN}(0,1).
	\end{equation} 
\end{proposition}
\begin{IEEEproof}
	Note that the fast fading for these channels is assumed to be independent. When the response of each reflective element is given, we have $\mathbb{E}[\Gamma_n h_n^{km}] = 0$ since $\mathbb{E}[h_n^{km}] = 0$. Besides, $\mathbb{E}[(\Gamma_n h_n^{km})^2] = \Gamma^2$ holds as $\mathbb{E}[(h_n^{km})^2] = 1$ and $|\Gamma_n| = \Gamma$. Therefore, according to the central limit theorem~\cite{HET-2013}, we have (\ref{central}) when $N$ is a sufficiently large number.
\end{IEEEproof}

\begin{remark}
	Proposition \ref{pro_central} implies that phase shifts of the RIS will not influence the distribution of the SNR when the number of reflective elements $N$ is sufficiently large. This observation shows that the expectation of the SNR will keep unchanged even when we select phase shifts of the RIS randomly.
\end{remark}

From this proposition, we can learn that the columns of $\bm{H}$ are zero-mean independent Gaussian vectors. Based on this, we can have the following proposition.

\begin{proposition}\label{upperbound}
	With the optimal power allocation, when the number of reflective elements is large but finite, the system capacity can be upper bounded by
	\begin{equation}\label{upper}
	C \leq  \sum \limits_{k} \log_2\left(1 + \frac{P \Lambda_k}{K\sigma^2}\bar{\beta}^k \Gamma^2MN\right).
	\end{equation}
\end{proposition}

\begin{IEEEproof}
The proof is given in Appendix \ref{proof_upperbound}.
\end{IEEEproof}

\begin{remark}
	The upper bound can be achieved when the eigenvalues of $\bm{GG}^H$ are the same. This can be obtained by optimizing the phase shifts of the RIS to maximize the effective rank of $\bm{G}$ if the problem has a feasible solution~\cite{PMG-2019}.
\end{remark}

It is worthwhile to point out that the system capacity will not increase infinitely. When the number of the reflective elements grows, the equivalent channel coefficient will decrease as well. In the following, we will give a proposition on the path loss $\bar{\beta}^k$ at the extreme case, i.e., $N \rightarrow \infty$. 

\begin{proposition}\label{pro_channel}
	Without loss of generality, we assume that the RIS is on the plane with $z = 0$, and we assume that the coordinates of the BS and user $k$ are $(-x_0,0,z_0)$ and $(x_0,0,z_k)$, respectively, as illustrated in Fig. \ref{illustration}. When $N \rightarrow \infty$, we have $\bar{\beta}_kN \approx \frac{2Az_0^3\lambda^2}{5\pi^2(z_0 + z_k)^5}$.
\end{proposition}
\begin{IEEEproof}
The detailed proof is given in Appendix \ref{pro_pro_channel}.	
\end{IEEEproof}

Based on the above two propositions, we have the following proposition.
\begin{proposition}\label{systembound}
When the number of reflective elements goes infinity, i.e., $N \rightarrow \infty$, the system capacity cannot exceed 
\begin{equation}
\tilde{C} = \sum \limits_{k} \log_2\left(1 + \frac{P\Lambda_k \Gamma^2 M}{K\sigma^2} \frac{2Az_0^3\lambda^2}{5\pi^2(z_0 + z_k)^5}\right).
\end{equation}
\end{proposition}

\section{Analysis on the Number of Reflective Elements}
\label{analysis}

In the following, we will discuss the effect of the number of reflective elements on the system sum-rate. We first give a proposition on the data rate for each user, and based on this, we will try to investigate how many reflective elements will be needed so that the sum-rate reach $\eta$ of the system capacity.  

\begin{proposition} \label{Sinr}
	The data rate for user $k$ can be rewritten by 
	\begin{equation}
	r_k = \mathbb{E}\left[\log_2\left(1 + \frac{P\Lambda_k}{K\sigma^2\left[\left(\bm{G}\bm{G}^H\right)^{-1}\right]_{k,k}}\right)\right].
	\end{equation}
\end{proposition}

\begin{IEEEproof}
	The proof is given in Appendix \ref{proof_sinr}.
\end{IEEEproof}

With this proposition, the sum rate can be written by $R = \sum \limits_{k} r_k$. Define $\epsilon$ as the ratio of the sum-rate to the system capacity with infinite reflective elements, i.e.,
\begin{equation}
\epsilon = R/\tilde{C}.
\end{equation}
Therefore, the problem can be written by
\begin{equation}\label{feas}
\min \limits_{N} N,~~s.t.~~ \epsilon \geq \eta.
\end{equation}

However, the aforementioned problem is hard to solve since the phase shifts of the reflective elements are not determined. In what follows, we first derive a lower bound of $\epsilon$, i.e., $\hat{\epsilon}$, to guarantee $\epsilon \geq \eta$, and then transform the original problem into
\begin{equation}\label{transformed}
\min \limits_{N} N,~~s.t.~~ \hat{\epsilon} \geq \eta.
\end{equation}
\subsection{Lower Bound of $\epsilon$}
The lower bound of the ratio of the sum-rate to the system capacity $\epsilon$ can be given by the proposition below.
\begin{proposition}
	$\epsilon$ can be lower bounded by
	\begin{equation}
	\hat{\epsilon} = \frac{\sum\limits_{k} \log_2\left(1 + \frac{P}{\sigma^2}\Lambda_k\Gamma^2\bar{\beta}^kN(\mu - 1)\right)}{\sum\limits_{k} \log_2\left(1 + \frac{P}{\sigma^2}\Lambda_k\Gamma^2\mu\frac{2Az_0^3\lambda^2}{5\pi^2(z_0 + z_k)^5}\right)},
	\end{equation}
	where $\mu = M/K$.
\end{proposition}
\begin{IEEEproof}
	According to the Jensen's inequality, we have
	\begin{equation}
	R \geq \sum\limits_{k} \log_2\left(1 + \frac{P\Lambda_k}{K\sigma^2 \mathbb{E}\left[\left[(\bm{GG}^H)^{-1}\right]_{k,k}\right]}\right).
	\end{equation}
	Here,
    \begin{equation}
    \left[(\bm{GG}^H)^{-1}\right]_{k,k} = \frac{\mathbb{E}[\text{Tr}\{(\bm{H}\bm{H}^H)^{-1}\}]}{K \bar{\beta}^k}.
    \end{equation}
    Note that $\frac{1}{N \Gamma^2} \bm{H}\bm{H}^H$ is a central Wishart matrix with $M$ degrees of freedom. Therefore, according to the results in~\cite{AS-2004}, we have
    \begin{equation}
    \mathbb{E}[\text{Tr}\{(\bm{H}\bm{H}^H)^{-1}\}] = \frac{K}{N\Gamma^2(M - K)}.
    \end{equation}
    
   Based on these, we have
   \begin{equation}
   \begin{aligned}
   \epsilon & \geq \frac{\sum\limits_{k} \log_2\left(1 + \frac{P\Lambda_k}{K\sigma^2}\Gamma^2\bar{\beta}^kN(M-K)\right)}{\sum\limits_{k} \log_2\left(1 + \frac{P\Lambda_k\Gamma^2M}{K\sigma^2}\frac{2Az_0^3\lambda^2}{5\pi^2(z_0 + z_k)^5}\right)},\\
   & = \frac{\sum\limits_{k} \log_2\left(1 + \frac{P}{\sigma^2}\Lambda_k\Gamma^2\bar{\beta}^kN(\mu-1)\right)}{\sum\limits_{k} \log_2\left(1 + \frac{P}{\sigma^2}\Lambda_k\Gamma^2\mu\frac{2Az_0^3\lambda^2}{5\pi^2(z_0 + z_k)^5}\right)}
   \end{aligned}
   \end{equation}
   which ends the proof.
\end{IEEEproof}

\begin{remark}\label{mu}
	From the lower bound of $\epsilon$, we can infer that a moderate number of antennas are required at the BS side to achieve an acceptable performance. Especially, when the number of antennas at the BS is equal to the number of users, we have $\mu = 1$, which leads to the lower bound of $\epsilon$ being 0.
\end{remark}

\subsection{Solution of Problem (\ref{transformed})}
To satisfy the constraint in (\ref{transformed}), we have
\begin{equation}
\prod \limits_{k} \frac{K + \frac{P}{\sigma^2} \Lambda_k \Gamma^2\bar{\beta}^kN(\mu - 1)}{\left(K + \frac{P}{\sigma^2}\Lambda_k \Gamma^2\mu \frac{2Az_0^3\lambda^2}{5\pi^2(z_0 + z_k)^5}\right)^{\eta}} \geq 1.
\end{equation}
We assume that the received SNR is high, i.e., $\frac{P}{\sigma^2} \Lambda_k \Gamma^2\bar{\beta}^kN(\mu - 1) \gg K$ and $\frac{P}{\sigma^2} \Lambda_k \Gamma^2\mu \gg K$. Thus, we have
\begin{equation}
\prod \limits_{k} \frac{\frac{P}{\sigma^2} \Lambda_k \Gamma^2\bar{\beta}^kN(\mu - 1)}{\left(\frac{P}{\sigma^2}\Lambda_k \Gamma^2\mu \frac{2Az_0^3\lambda^2}{5\pi^2(z_0 + z_k)^5}\right)^{\eta}} \geq 1.
\end{equation}
In other words,
\begin{equation}\label{number_of_element}
N \geq  \left(\prod \limits_{k} \frac{\left(\frac{P}{\sigma^2}\Lambda_k \Gamma^2\mu \frac{2Az_0^3\lambda^2}{5\pi^2(z_0 + z_k)^5}\right)^{\eta}}{\frac{P}{\sigma^2} \Lambda_k \Gamma^2\bar{\beta}^k(\mu - 1)}\right)^{1/K}.
\end{equation}

	
	


\section{Simulation Results}
\label{simulation}


\begin{figure}[!t]
	\centering
	\includegraphics[width=2.8in]{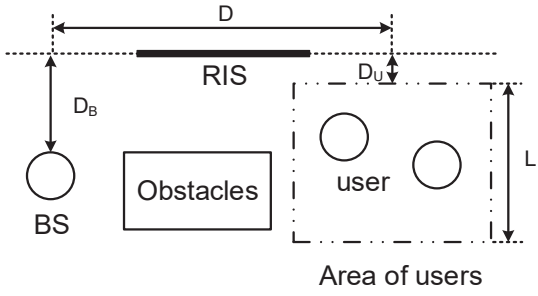}
	\vspace{-2mm}
	\caption{Simulation layout for the RIS-based multi-user cellular network (top view).}
	\vspace{-2mm}
	\label{layout}
\end{figure}

In this section, we verify the derivation of the achievable asymptotic capacity and evaluate the impact of the number of reflective elements on the data rate in the RIS assisted multi-user communications, the layout of which is given in Fig. \ref{layout}. The parameters are selected according to 3GPP standard~\cite{3GPP-2018} and existing works \cite{HBLZ-2020}. The height of the BS is 25m and the distance between the RIS and the BS $D_B = 100$m. We set the number of users $K = 5$. The users are uniformly located in a square area whose side length is set as $L = 100$m. The distance between the RIS and the closest side of the square area is $D_u = 10$m, and the horizontal distance between the BS and the center of the square area is $D = 100$m. The center of the RIS is located at the middle between the BS and the square area with the height being 25m. The working frequency of the RIS is set as $f = 5.9$GHz and reflection amplitude is assumed to be $\Gamma = 1$. The length and width of a reflective element are set the same, i.e., $a = b = 0.02$m. Transmit power is set as $P = 46$ dBm, noise power is set as $\sigma^2 = -96$ dBm, and antenna gain is $A = 0$dB. All numeral results are obtained by 100 Monte Carlo simulations.

\begin{figure}[!t]
	\centering
	\includegraphics[width=3.2in]{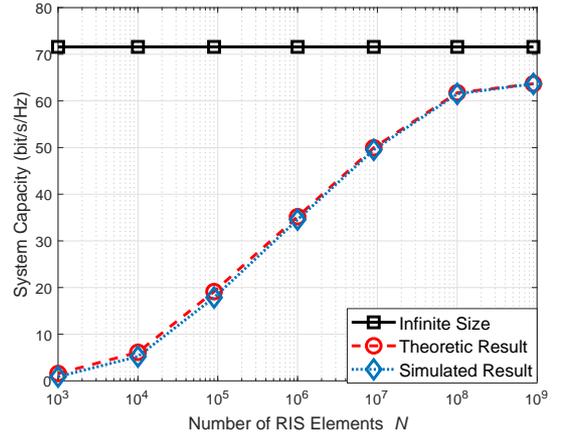}
	\vspace{-2mm}
	\caption{System capacity vs. Number of reflective elements $N$ with the number of antennas $M = 10$.}
	\vspace{-4mm}
	\label{data}
\end{figure}

In Fig.~\ref{data}, we plot the system capacity vs. the number of reflective elements $N$ with the number of antennas $M = 10$. From this figure, we can observe that the theoretic result is a tight approximation of the simulated one, and the gap will decrease as the number of reflective elements increases because it can make the singular values of the channel matrix more equal. Moreover, we can observe that the capacity per user will increase as the number of reflective elements $N$ increases first and then become saturated, as suggested by Proposition \ref{systembound}. We can observe that  the system capacity with an infinite size of the RIS can reach 71.5 bit/s/Hz for 5 users.

\begin{figure}[!t]
	\centering
	\includegraphics[width=3.2in]{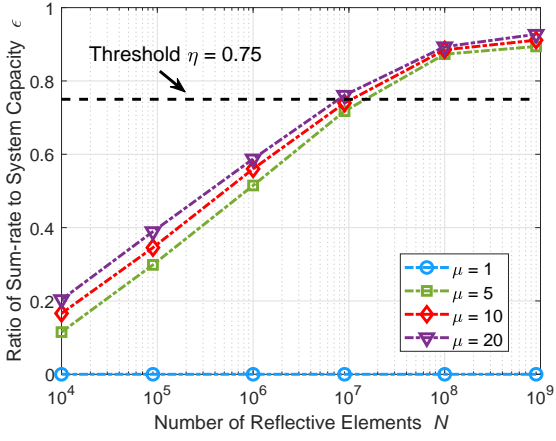}
	\vspace{-2mm}
	\caption{Ratio of sum-rate to system capacity $\hat{\epsilon}$ vs. Number of reflective Elements $N$.}
	\vspace{-4mm}
	\label{quantization}
\end{figure}

In Fig.~\ref{quantization}, we plot the ratio of sum-rate to system capacity $\hat{\epsilon}$ vs. the number of reflective elements $N$ with threshold $\eta = 0.75$ for different values of $\mu$. From this figure, we can observe that the ratio $\hat{\epsilon}$ first increases linearly with $\log N$ and then saturates, which suggests that it is more cost-effective to operate the RIS on the linear growth zone, i.e., $N < 10^7$ . We can also observe that it requires the number of reflective elements $N = 8\times10^6$ to achieve 75\% of the system capacity with $\mu = 20$. In other words, the side length of the RIS should be 56.7m. Besides, we can observe that the required number of reflective elements will decrease with a higher value of $\mu$. This implies that the size of the RIS can be reduced with more antennas at the BS. Therefore, there exists an trade-off between the number of reflective elements and the number of antennas at the BS, and we can reduce the total cost by optimizing the size of the RIS. Moreover, we can find that the ratio remains 0 when $\mu = 1$, i.e., the number of antennas at the BS is the same as the number of users. This implies that the antennas at the BS should be larger than the number of users, which is consistent with remark \ref{mu}. 

\section{Conclusions}
\label{sec:conclusion}
In this letter, we have investigated RIS assisted downlink multi-user communications and have derived the corresponding asymptotic capacity. Based on the derived capacity, we also have discussed the impact of the number of the reflective elements on the system sum-rate. Particularly, we have obtained a requirement on the RIS size to achieve an acceptable data rate. From the analysis and simulation, given the location of the RIS, we can draw the following conclusions: 1) The system capacity cannot increase infinitely as the RIS size grows, but rather is upper bounded; 2) To achieve a data rate threshold, the required number of reflective elements will decrease with more antennas at the BS.

\begin{appendices}
	
	\section{Proof of Proposition \ref{upperbound}}\label{proof_upperbound}
	According to Jensen's inequality, we have
	\begin{equation}
	\begin{aligned}
	C &\leq \sum \limits_{k} \log_2 \left(1 + \frac{P \Lambda_k}{K\sigma^2}\mathbb{E}\left[[\bm{GG}^H]_{k,k}\right]\right)\\
	& = \sum \limits_{k} \log_2 \left(1 + \frac{P \Lambda_k}{K\sigma^2}\frac{\bar{\beta^k}}{K}\mathbb{E}[\text{Tr}\{\bm{H}\bm{H}^H\}]\right).
	\end{aligned}
	\end{equation}
	
	Based on Proposition \ref{pro_central}, $\frac{1}{N \Gamma^2}\bm{H}\bm{H}^H \sim \mathcal{W}_{K}(M,\bm{I}_M)$ is a central Wishart matrix with $M$ degrees of freedom where $M \geq K$. According to the results in~\cite{AS-2004}, we have 
	\begin{equation}
	\mathbb{E}[\text{Tr}\{\bm{H}\bm{H}^H\}] = MNK\Gamma^2.
	\end{equation}
	Therefore, we can have (\ref{upper}).
	
	\section{Proof of Proposition \ref{pro_channel}}\label{pro_pro_channel}

	Based on the coordinates given in Fig.~\ref{illustration}, we have
	\begin{equation}
	\beta_{n}^k = \frac{Aabz_0^3\lambda^2}{64\pi^3((x - x_0)^2 + y^2 + z_k^2)((x + x_0)^2 + y^2 + z_0^2)^{5/2}}.
	\end{equation}
	Therefore, we have $\bar{\beta}^k N = \sum\limits_{n = 1}^N \beta_{n}^k$.
	
 	When $N \rightarrow \infty$, in the most area of the RIS, the distance from the reflective element and the center between the BS and user $k$ is much longer than that between the BS and user $k$. Therefore, we can assume that the distances from the reflective element to the BS and the user are almost the same, and thus have the approximations below:
	\begin{equation}
	\begin{aligned}
	((x - x_0)^2 + y^2 + z_k^2) &\approx ((x + x_0)^2 + y^2 + z_0^2)\\
	& \approx (x^2 + y^2 + (z_0 + z_k)^2/4).
	\end{aligned}
	\end{equation} 
	Define the area of the RIS as $\Phi$. The received energy of the signals reflected by reflective elements in this area can be expressed by
	\begin{equation}
	\begin{aligned}
	&\sum\limits_{(x,y) \in \Phi} \beta_{n}^k \approx \int \limits_{(x,y) \in \Phi} \frac{Az_0^3\lambda^2}{64 \pi^3(x^2+y^2+(z_0 + z_k)^2/4)^{7/2}} dxdy\\
	&= \frac{Az_0^3\lambda^2}{32\pi^2} \int \limits_{0}^{+\infty} \frac{1}{(\rho^2 + (z_0 + z_k)^2/4)^{7/2}} \rho d \rho  = \frac{2Az_0^3\lambda^2}{5\pi^2(z_0 + z_k)^5},
	\end{aligned}
	\end{equation}
	where $\rho^2 = x^2 + y^2$.
	
	\section{Proof of Proposition \ref{Sinr}}\label{proof_sinr}
		According to the definition in (\ref{sinr}), since $\frac{P\Lambda_k}{K}$ is a constant, we can derive $(\bm{g}_k\bm{w}_k)(\bm{g}_k\bm{w}_k)^{H}$ as follows:
		\begin{equation}
		(\bm{g}_k\bm{w}_k)(\bm{g}_k\bm{w}_k)^{H} = \bm{g}_k \frac{\bm{v}_k}{\|\bm{v}_k\|_F}\frac{\bm{v}^H_k}{\|\bm{v}_k\|_F}\bm{g}^H_k \overset{(a)}{=} \frac{1}{\|\bm{v}_k\|_F^2}.
		\end{equation}
		Here, (a) can be achieved by
		\begin{equation}
		\begin{array}{ll}
		\bm{g}_k \bm{v}_k\bm{v}^H_k\bm{g}^H_k
		&\hspace{-2mm}= \left[\bm{G}\bm{G}^{H}(\bm{G}\bm{G}^{H})^{-1}\right]_{k,k} \left[\bm{G}\bm{G}^{H}(\bm{G}\bm{G}^{H})^{-1}\right]^{H}_{k,k}\\
		&\hspace{-2mm} = 1.
		\end{array}
		\end{equation}
		Note that $\|\bm{v}_k\|_F^2 = \bm{v}^{H}_k\bm{v}_k$, we have
		\begin{equation}
		\begin{array}{ll}
		\bm{v}^{H}_k\bm{v}_k &\hspace{-2mm}= \left[\bm{V}^H\bm{V}\right]_{k,k}  = \left[(\bm{G}\bm{G}^H)^{-1}\bm{G}\bm{G}^H(\bm{G}\bm{G}^H)^{-1}\right]_{k,k}\\
		& \hspace{-2mm}= \left[(\bm{G}\bm{G}^H)^{-1}\right]_{k,k}.
		\end{array}
		\end{equation}

\end{appendices}


\vspace{-3mm}
\begin{thebibliography}{40}
	
\bibitem{M-2019}	
M. Renzo, et al,  ``Smart Radio Environments Empowered by Reconfigurable AI Meta-surfaces: An Idea Whose Time Has Come," \emph{EURASIP J. Wireless Commun. Netw.}, vol. 2019, no. 129, pp. 1-20, May 2019.

\bibitem{MHLKZG}
M. A. Elmossallamy, H. Zhang, L. Song, K. Seddik, Z. Han, and G. Y. Li, ``Reconfigurable Intelligent Surfaces for Wireless Communications: Principles, Challenges, and Opportunities," \emph{IEEE Trans. Cognitive Commun. Netw.}, vol. 6, no. 3, pp. 990-1002, Sep. 2020.

\bibitem{BHLYZH}
B. Di, H. Zhang, L. Song, Y. Li, Z. Han, and H. V. Poor, ``Hybrid Beamforming for Reconfigurable Intelligent Surface based Multi-user Communications: Achievable Rates with Limited Discrete Phase Shifts," \emph{IEEE J. Sel. Areas Commun.}, vol. 38, no. 8, pp. 1809-1822, Aug. 2020.
\bibitem{YCZDYJ-2020}
Y. Cao, C. Yong, Z. Xiong, D. Niyato, Y. Xiao, and J. Zhao, ``Reconfigurable Intelligent Surface for MISO Systems with Proportional Rate Constraints", in \emph{Proc. IEEE ICC}, London, UK, Jun. 2020.

\bibitem{HBLZ-2020}
H. Zhang, B. Di, L. Song, and Z. Han, ``Reconfigurable Intelligent Surfaces assisted Communications with Limited Phase Shifts: How Many Phase Shifts Are Enough?" \emph{IEEE Trans. Veh. Technol.}, vol. 69, no. 4, pp. 4498-4502, Apr. 2020.

\bibitem{OEE-2020}
W. Tang, M. Chen, X. Chen, J. Dai, Y. Han, M. Renzo, Y. Zeng, S. Jin, Q. Cheng, and T. Cui, ``Wireless Communications with Reconfigurable Intelligent Surface: Path Loss Modeling and Experimental Measurement", \emph{IEEE Trans. Wireless Commun.}, to be published. 

\bibitem{DP-2005}
D. Tse and P. Viswanath, \emph{Fundamentals of Wireless Communications}. Cambridge Univ. Press, Cambridge, U.K., 2005.

\bibitem{HET-2013}
H. Q. Ngo, E. G. Larsson, and T. L. Marzetta, ``Energy and Spectral Efficiency of Very Large Multiuser MIMO Systems," \emph{IEEE Trans. Commun.}, vol. 61, no. 4, pp. 1436–1449, Apr. 2013.

\bibitem{AS-2004}
A. M. Tulino and S. Verd{\'u}, ``Random Matrix Theory and Wireless Communications", \emph{Foundations Trends Commun. Inf. Theory}, vol. 1, no. 1, pp. 1-182, Jun. 2004.

\bibitem{PMG-2019}
M. A. Elmossallamy, H. Zhang, R. Sultan, K. Seddik, L. Song, G. Y. Li, and Z. Han, ``On Spatial Multiplexing Using Reconfigurable Intelligent Surfaces," \emph{IEEE Wireless Commun. Lett.}, to be published.

\bibitem{3GPP-2018}
3GPP TR 38.901, ``Study on Channel Model for Frequencies from 0.5 to 100 GHz (Release 14)," Jan.~2018.





\end{thebibliography}
\end{document}